\documentstyle[aps,preprint]{revtex} 
\begin{document}
\draft 
\title{Ion-Chain Dynamics in Polymer Electrolytes}

\author{L. D. Carlos$^{(1)}$ and A. L. L. Videira$^{(2)}$}
\address{$^{(1)}$ Departamento de F\'{\i}sica, Universidade de Aveiro,
3800 Aveiro, Portugal}

\address{$^{(2)}$ Departamento de F\'{\i}sica, Universidade de \'Evora,
Ap. 94, 7001 \'Evora Codex, Portugal; Centro de Matem\'atica e
Aplica\c c\~oes Fundamentais, Universidade de Lisboa, Av. Prof. Gama 
Pinto, 2, 1699 Lisboa, Portugal; and Academia Militar, Pa\c co da Rainha,
29, 1198 Lisboa, Portugal}
\date{\today}
\maketitle
\begin{abstract}
Representing polyether-salt systems by chains of interacting coordination
shells, defined by the cation and by its nearest ligands, we derive the
interaction potential between closest shells -- the inter-shells 
potential --
in terms of two-electron polarization effects. Values are presented for
monovalent-based crystalline poly(ethylene oxide), PEO, electrolytes. For 
the eutectic composition $\text{PEO}_{12} \text{EuBr}_3$, the inter-shells 
energy is evaluated also by relating the empirical value of the 
nearest-ligands local-field potential with the variation of the 
$\text{Eu}^{3+}$ concentration. Both methods give the same results.  
\end{abstract}
\renewcommand{\baselinestretch}{1.656} 
\narrowtext
In spite of extensive work performed over the past two decades 
on polymer electrolytes modified by monovalent and multivalent 
cations, the interpretation of the abundant empirical results 
has been hindered by the lack of detailed knowledge of their 
ion-chain structures. Because polymer electrolytes present 
certain liquid-like features, with the polymer acting as an 
''immobile'' solvent \cite{Armand89}, the structural properties 
of these materials are predominantly determined by their local 
configuration in the vicinity of the cation, and depend on the 
number of complexed cations, that is, their structural features 
change with the number $n$ of ether-oxygens in the polymer chain 
per cation. For crystalline monovalent-based poly(ethylene oxide), 
PEO, electrolytes, the ion-chain local configuration has been 
depicted as forming oxygen-lined helical turns, with the cation 
located within the polymer cavity \cite{Armand79}, a configuration 
which has been confirmed by single crystal X-ray diffraction 
\cite{Chatani,Bruce} for a few salt-rich complexes. Following this 
idea, in this Letter, polyether-salt systems are represented
by chains of interacting coordination shells, each of which is 
defined by a cation at its center and by its nearest ligands
(ether-oxygens and, eventually, anions) at its boundary. In this 
model, the ion-chain local dynamics -- which changes with the number 
of interacting shells, that is, with $n$ -- results jointly from 
each shell's cation-nearest ligands interaction -- the intra-shell 
interaction -- and from the interaction between a given coordination 
shell and all the others in its immediate vicinity -- the inter-shells 
interaction. (For crystalline electrolytes, these neighborhoods 
may be approximated to their unit cells)

We here derive an expression for the inter-shells potential, 
$V_{ss}$, between all nearest-shells, whose ligands are separated 
by an average inter-shells distance $R_{ss}$, in terms of two-electron 
polarization effects \cite{Judd77}, mediated by the ligands' valence
electrons (mainly the 2{\it p} oxygen electrons) of each interacting 
shell. For a given shell, we consider the polarization of one of 
its ligands located at ${\vec {\text R}}$ from the cation 
(taken as the origin) by one of the valence electrons of the 
neighboring shells, at ${\vec {\text r}}_1$; one of the remaining 
electrons at ${\vec {\text r}}_2$ interacts with the multipole moments 
induced in the polarized ligand. Thus, we write for the interaction 
between a given coordination shell and all the others closest to it

\begin{eqnarray}
V_{ss}^n&=&\sum_{ab\alpha} \sum_{k\beta m} F_n (abk)
\frac{R^a C_{a+n,\alpha +m}^{\ast} ({\hat r}_1)} 
{(r_1)^{a+n+1}} \frac{R^b C_{b+n,\beta +m} 
({\hat r}_2)} {(r_2)^{b+n+1}} \nonumber \\
&\times &\left( \begin{array}{ccc} b+n & a+n & k \\ 
-\beta -m & \alpha +m & -\alpha +\beta \end{array} \right) 
C_{k,\beta -\alpha }^{*}(\hat R) \, ,
\end{eqnarray}
where $C_{\ell,q}=(4\pi /2\ell+1)^{1/2} Y_{\ell,q}$ are spherical 
tensors, $F_n(abk)$ -- expressed in terms of 3-j symbols ( ) and 
of 6-j symbols \{ \} -- is $F_n(abk)=(-\alpha_n e^2) (-1)^{n-m+\beta} 
(2k+1) \left( \begin{array}{ccc} a & b & k \\ 0 & 0 & 0 \end{array} 
\right) \left\{ \begin{array}{ccc} a & a+n & n \\ b+n & b & k 
\end{array} \right\} [(2a+2n+1)(2b+2n+1)]^{1/2} \left[ \left( 
\begin{array}{c} {2a+2n} \\ {2a} \end{array} \right) \left( 
\begin{array}{c} {2b+2n} \\ {2b} \end{array} \right) \right]^{1/2}$,
and $\alpha _n$ is the nth multipole polarizability of the ligands. 
Taking only the dipolar contribution, assuming spherical symmetry 
in the space of the ligands valence electrons ($k$=0), 
$r_1=r_2={\text R}_{ss}+{\text R}$, and performing the sum over 
these electrons and over all the $N$ ligands of valence Z$_i$ at 
${\vec {\text R}_i}$, we write the inter-shells interaction as

\begin{equation}
V_{ss}={\frac 1{{2}}} \sum_i^N \alpha_i e^2 Z_i \sum_{a=0}^\infty
(a+1)\ {\frac{R_i^{2a}}{{{(R_i+R_{ss})}^{2a+4}}}} \ . 
\label{vss1}
\end{equation}
Considering only the first two terms of the expansion 
($R_{ss}\gg R_i$), and considering that all the ligands of each 
shell are at the same average distance $\bar {\text R}$ from the 
cation, we have 

\begin{equation}
V_{ss}={\frac 1{{2}}}\sum_i^N \alpha_i e^2 Z_i \left( \frac 1{(\bar {R}+
R_{ss})^4} + \frac {2\bar {R}^2} {(\bar {R}+R_{ss})^6}\right) \, .
\label{vss2} 
\end{equation}

The application of Eq.\ (\ref{vss2}) to polyether-salt systems 
requires the knowledge of the dipole polarizability $\alpha_i$, 
the number of ligands $N$, the cation-ligands average distance 
$\bar {\text R}$, and the distance between closest shells 
${\text R}_{ss}$. These parameters have been determined, through 
single crystal X-ray diffraction, for a few crystalline 
monovalent-based PEO electrolytes \cite{Chatani,Bruce}, and the 
corresponding inter-shells potentials calculated from 
Eq.\ (\ref{vss2}) are given in Table\ \ref{tab1}. For 
$\text{PEO}_3 \text{NaI}$, Wright \cite {Wright} estimates the total 
interaction energy between the $\text{Na}^+$ and $\text{I}^-$ ions, 
and the 3EO polymer segments of neighboring helices as being 
$\varepsilon \text{(3EO-NaI-inter)}$ $\approx \, 20-30$ 
kJ (mol of 3EO units)$^{-1}$ ($\approx 1700-2500 \text{cm}^{-1}$), 
a value which is of the same order of magnitude as the one calculated 
in Table \ref{tab1}.

For non-crystalline divalent polyether-salt complexes, the only 
local structural information is provided by extended X-ray 
absorption fine-structure measurements (EXAFS). These results, which 
lie within a spherical domain of up to $6\, \text{\AA}$ of the target 
cation, besides giving the number $N$ of ether-oxygens of the first 
coordination shell (the anions being further away from the cation), 
also furnish the average distances $\bar {\text R}$ (some of which are 
referred to in Table III of Ref. \cite{Carlos94b}), but only a lower 
limit for $\text {R}_{ss}+ \bar {\text R}>6 \, \text{\AA}$. The 
corresponding upper limits for $V_{ss}$ are, for example, 
$V_{ss}<600\, \text{cm}^{-1}$ in $\text{PEO}_n \text{ZnI}_2$ 
($30\geq n\geq 4$), and $V_{ss}<900 \, \text {cm}^{-1}$ in 
$\text{PEO}_n \text{ZnBr}_2$ ($20\geq n\geq 6$). 

Very little is known about the ion-chain structure of trivalent-based
polymer electrolytes. In particular, there are very few results 
for $\bar {\text R}$ \cite{Carlos94b,Carlos95a} (in this case, 
the anions are also supposed to be further away from the coordination 
cation), and none for $\text {R}_{ss}$. Here, we evaluate 
$\text {R}_{ss}$ in lantahanide-based electrolytes for the 
concentration for which the interaction between nearest shells 
does not yet modify the internal dynamics of each shell. That is, 
we determine the minimum inter-shells distance, 
$\text{R}_{ss}= \text{R}_0$, for which $V_{ss}$ does not yet 
alter $\bar {\text R}$. For this, following Morrison \cite{Morrison80}, 
we write the intra-shell potential $V_s$, also due to two-electron 
polarization effects, as    

\begin{equation}
V_s=-{\frac 1{{2}}} \sum_i \alpha_i e^2 Z_i \left( 
\frac 1{\bar {R}^4} + \frac{2\rho ^2} {\bar {R}^6} \right) \ ,
\label{vs} 
\end{equation}
where $\rho$ is an ion-dependent, host-independent quantity, which
corrects the Hartree-Fock expectation values of the even powers of 
the 4{\it f} electrons radial distances. Imposing a balance
between the repulsive $V_{ss}$, Eq.\ (\ref{vss2}), and the 
attractive $V_s$, Eq.\ (\ref{vs}), expressed by 
$(dV_{ss}/dR)_{R=R_0}=(dV_s/dR)_{R=\bar {R}}$, we get

\begin{equation}
(\bar R+R_0)^7({\bar R}^2+3\rho ^2)+{\bar R}^7R_0^2+2{\bar R}^8R_0+
4{\bar R}^9=0 \ ,
\label{R0} 
\end{equation}
from which, knowing $\bar {\text R}$, the minimum inter-shells 
distance $R_0$ may be determined, and, therefore, from 
Eq.\ (\ref{vss2}), also the associated lanthanide inter-shells 
potential. $\bar {\text R}$ has been determined only for 
$\text{PEO}_n \text{EuBr}_3$, for which the Eu$^{3+}$ local 
surroundings were investigated by means of the observed 
$^5D_{0-1} \rightarrow {^7}F_{0-4}$ emission lines. Taking the 
resulting value $\bar {\text R}=2.4 \ \text{\AA}$ for 
$80\geq n\geq 8$ \cite{Carlos94b,Carlos95a}, and 
$\rho^2=0.1666\ \text{\AA}^2$ \cite{Morrison80} into Eq.\ (\ref {R0}), 
we get $\text {R}_0=5.3\ \text{\AA}$. Table\ \ref{tab2} presents
the results for the inter-shells potential for a number 
of europium-nearest oxygens between 8 and 12, a value in accordance 
with both the determined $C_{2v}$ local symmetry 
group \cite {Carlos95a,Carlos94a}, generally associated with a 
coordination number around 8, and with a recent 
result on $\text{La}^{3+}$-based PEO electrolytes \cite{Bernson}, 
which indicates between 9 and 10 oxygens surrounding the lanthanum 
ion. 

Next, for the $\text{PEO}_n \text{EuBr}_3$ electrolytes, we 
represent the effect of the increase of the Eu$^{3+}$ concentration 
from $n=80$ to $n=5$ on the interaction potential between the 
europium and its surrounding oxygens -- the intra-shell potential. 
As usual, we take this potential as the local-field interaction 
$V_{LF}=\sum_i^6 \sum_{kq} B_{k,q} C_{k,q}(\theta _i,\phi_i)$, with 
the spherical coordinates ($r_i,\theta _i,\phi _i$) fixing the 
location of the $\text{Eu}^{3+}$ ith 4{\it f} electron, and 
$B_{k,q}$ ($k=0,2,4,6$; $-k\leq q\leq k$) being phenomenological 
parameters encompassing all the dynamics of $\text {Eu}^{3+}$ in 
the complex. Then, relating the dynamical part of $V_{LF}$ -- 
the empirically determined $B_{k,q}$ -- with the interaction energy 
between nearest shells -- the inter-shells potential -- we obtain, 
by an independent method, and for the $\text {PEO-EuBr}_3$ stablest
concentration $n\approx 12$, the inter-shells interaction energy 
$\varepsilon$.

The emission spectra of these electrolytes present a series
of sharp lines assigned to transitions between the Stark components 
of the $^5D_{0,1}$ and $^7F_{0-4}$ 
levels \cite{Carlos95a,Carlos94a,Carlos95b}, of which the only 
transitions considered here are the dominant 
electric-dipole $^5D_0 \rightarrow {^7}F_2$ and the magnetic-dipole 
$^5D_{0}\rightarrow {^7}F_1$ lines. The eigenvalues of the 
matrix elements of the local-field perturbation for the 
$^7F_{1,2}$ Stark sublevels are expressed in terms of the 
non-zero $B_{2,q}$ and $B_{4,q}$ parameters (the $k=0$ term not 
being included, as it merely shifts the energetic 
configuration as a whole), and are fixed by the best fitting 
between the observed Stark energies and the calculated eigenvalues 
\cite{Carlos95a,Carlos94a,Carlos95b}. Considering -- as in Newman's
superposition model \cite{Newman} -- that only the nearest ligands 
contribute significantly to the phenomenological intra-shell 
potential $V_{LF}$, the most relevant parameters are the 
$B_{4,q}$ (as the $k=2$ terms relate to contributions beyond the 
first coordination shell). Since, for lanthanides complexes, the 
$B_{4,q}$ predicted by the point-charge model -- in which 
the dynamical parameters depend on the angular distribution of the 
4{\it f} electrons, on the ion-first neighbors distance, on the 
ligands valence, and on the number of cation-nearest neighbors -- 
are in good agreement with the corresponding empirical values, 
in the case of $\text{PEO}_n \text{EuBr}_3$ electrolytes, the effect 
of the increase of $n$ on the intra-shell potential is represented 
expressing these $B_{4,q}$ in terms of that model. These parameters 
change as the $\text{Eu}^{3+}$ concentration increases from 
$n=80$ to $n=5$ \cite{Carlos94a,Carlos95b}. Since, in this concentration 
range, neither $\bar {\text R}$ nor $N$ vary, the change of the 
phenomenological intra-shell potential $V_{LF}$ must result 
from a modification of the ligands valence. The connection between 
the electrolyte {\it global} behavior -- expressed by the ion's 
concentration increase -- and the {\it local} interaction within 
each coordination shell -- expressed by the sum of all the 
$B_{4,q}$ -- must, therefore, be given by the changes in the 
ligands valence. In the model presented here of $\text{PEO-EuBr}_3$ 
electrolytes as chains of interacting luminescent shells -- the 
number of which increases with the amount of 
complexed-$\text{Eu}^{3+}$ --, the intra-shell potential 
$V_{LF}$ (and, therefore, the ligands valence) must reflect the
increase of the interaction between nearest shells with the decrease 
of the oxygen-cation ratio. For a europium concentration increase 
from $n=80$ to $n=32$, the coordination shells are spaced so far apart 
that they do not interact with each other. The existence of free PEO 
in this concentration range \cite{Carlos95d} indicates the coexistence 
of complexed and non-complexed ether-oxygens in the electrolytes. In 
the range $32\geq n\geq 8$, the number of luminescent shells present 
in the electrolytes is such that the distance between nearest shells 
decreases in such a way that the induced ligands valence charge of 
each shell also becomes smaller. This, in turn, is associated with 
a corresponding diminution of the intra-shell potential. Hence, in 
this concentration range, the inter-shells interaction begins to become 
relevant without yet altering the average $\text{Eu}^{3+}$-ligand 
oxygens distance. The constancy, between $n=80 $ and $n=8$ of the 
mean radius $\bar {\text R}$ of each 
coordination shell suggests that, as the concentration of 
$\text{Eu}^{3+}$ increases, there should not be any significant 
modification of the internal dynamics of the shells. For $n<8$, the
electrostatic repulsion between nearest shells increases, their mutual
interaction decreases, and the electrolytes are not found any more in their 
stablest phase -- the eutectic concentration $n\approx 12$. This behavior 
of the intra-shell dynamics with the increase of $n$ shows precisely the 
same trend as the observed dependence with $n$ of the ratio between 
the $^5D_0\rightarrow {^7}F_2$ and $^5D_0\rightarrow {^7}F_1$ average 
intensities, $I_{0-2}/I_{0-1}$, which slowly diminishes between 
$80\geq n\geq 12$, attains a minimum for $n\approx12$, and sharply 
increases for $n<12$ \cite{Carlos95b}. As it is well known that the 
relative intensities of the above two transitions (the only 
ones being considered here) are hypersensitive to the nature of the 
$\text{Eu}^{3+}$-ligand surroundings, the intra-shell potential $V_{LF}$ 
must, somehow, incorporate the ratio $I_{0-2}/I_{0-1}$. We define the 
strength of this local-field potential as $s_{LF}^2\equiv (2k+1)s_{k=4}^2$,
where $s_{k=4}^2=(2k+1)^{-1}\sum_q(B_{4,q})^2$ is the corresponding 
Leavitt's quadratic rotational invariant \cite{Leavitt}. We then 
propose that, in terms of this strength, the phenomenological 
potential is related to the $\text{Eu}^{3+}$ concentration by

\begin{equation}
s_{LF}\equiv \left( {\sum_q} \left(B_{4,q} \right)^2 \right)^{1/2}
= \varepsilon \left(1-e^{1-n/n_0} \right)^2 
\label{str}
\end{equation}
where $\varepsilon$ is the interaction energy between a given 
coordination shell and all the other shells closest to it, corresponding 
to the $\text{Eu}^{3+}$ stablest concentration $n_0$. (Here, the 
radial dependence of the $B_{4,q}$ is manifested on the variation 
of the cation concentration $n$ with the distance between nearest 
coordination shells.) Inserting into Eq.\ (\ref{str}) the value for 
$s_{LF}$ obtained from the phenomenological 
$B_{4,q}$ \cite{Carlos94a,Carlos95b}, we evaluate $\varepsilon $ and 
$n_0$ by the least squares fit (Fig. \ref{fig1}). The results 
obtained were $\varepsilon =554.2\ \text{cm}^{-1}$ and $n_0=12.9$. 
This value for the $\text{Eu}^{3+}$ stablest concentration 
is precisely the one given by the morphological results (DSC, XRD, 
etc.) \cite {Carlos95d}, thus bearing out the relation proposed in 
Eq.\ (\ref{str}). 

In summary, representing polyether-salt systems by chains of 
interacting coordination shells, defined by the cation and its 
nearest ligands, we derive the interaction potential between a 
given coordination shell and all the others in its immediate 
surroundings in terms of two-electron polarization effects. The 
values obtained for monovalent-based crystalline PEO electrolytes 
are given in Table \ref{tab1}. For non-crystalline divalent 
electrolytes we are only able to estimate an upper limit for 
the inter-shells potential. For $\text{PEO-EuBr}_3$ 
electrolytes, we deal with this inter-shells potential in two 
separate ways, one involving the same two-electron polarization 
effects, and the other relating the empirical value of the 
nearest-ligands local-field potential with the variation of 
$\text{Eu}^{3+}$ concentration. The inter-shells interaction energy, 
$\varepsilon $, calculated by the the two-electron 
polarization potential is $\varepsilon=520.3\ \text{cm}^{-1}$ for 
10 nearest oxygens, and is $\varepsilon=572.4\ \text{cm}^{-1}$, for 
11 oxygens. The value of $\varepsilon $ evaluated by the 
phenomenological local-field potential is 
$\varepsilon =554.2\ \text{cm}^{-1}$ for the 
$\text{PEO-EuBr}^{3+}$ stablest concentration $n_0 \approx 12$. The 
agreement between the two independent methods indicates that the 
concentration for which the inter-shells interaction does not yet 
alter the intra-shell dynamics is the stablest concentration 
($n\approx 12$). In addition, this agreement also suggests that 
the screening effects of the non-complexed oxygens, included in 
the inter-shells potential, Eq.\ (\ref{vss2}), are well accounted 
for by the dipole polarization, by the number of ligands, and by 
their valence. On the other hand, although, as previously referred to 
\cite{Carlos94b}, the dependence of the intra-shell potential with the 
$\text{Eu}^{3+}$-ligands distance, Eq.\ (\ref{vs}), is 
essentially correct, the screening effects of the 5{\it s} and 5{\it p} 
shells is eventually over estimated. This may be due to the distance 
between two nearest shells being more than twice the 
europium-oxygens distance.

This research was co-financed by FEDER and JNICT, STRIDE 
program (Portugal), STRDB/C/CTM/628/92. 

\begin{figure}
\caption
{Dependence of the strength of the nearest-ligands local-field 
interaction $s_{LF}$ ($\text{cm}^{-1}$) on O/Eu ratio ($n$) for 
$\text{PEO}_n \text{EuBr}_3$ (n=5,8,12,16,20,24,28,32,80). Fitting 
parameter $\Delta=40 \ \text{cm}^{-1}$.}
\label{fig1}
\end{figure}

\begin{table}
\caption{Coordination number $N$, nearest cation-ligands average 
distance $\bar {\text R}\, \text{(\AA)}$, inter-shells distance 
${\text R}_{ss} \, \text{(\AA)}$, and inter-shells potential 
$V_{ss}$ $\text{cm}^{-1}$), for a few monovalent-based PEO 
electrolytes.}
\label{tab1}
\begin{tabular}{ccccc}
& $N$ & $\bar {\text R}$ & ${\text R}_{ss}$ & $V_{ss}$ \\ 
\tableline                        
$\text {PEO}_3 \text{NaI}$ & 5 \tablenotemark[1] & 2.71 & 2.98 
& 1453.23 \\
$\text {PEO}_3 \text{NaSCN}$ & 6 \tablenotemark[2] & 2.47 & 5.69 
& 415.46 \\
$\text {PEO}_3 \text{NaClO}_4$ & 5 \tablenotemark[3] & 2.50 & 3.35 
& 915.85 \\
$\text {PEO}_3 \text{LiCF}_3 \text{SO}_3$ & 5 \tablenotemark[3] & 2.10 
& 4.41 & 526.65 \\
$\text {PEO}_4 \text{KSCN}$ & 7 \tablenotemark[4] & 2.87 & 2.34 
& 3742.49 \\ 
$\text {PEO}_4 \text{NH}_4 \text{SCN}$ & 7 \tablenotemark[4] & 2.96 
& 2.31 & 3638.94 \\
\end{tabular}
\tablenotetext[1]{3 oxygen and 2 iodine ligands, $\alpha(O^{2-})=1.349 \
\text{\AA}^3$ and $\alpha(I^{-})=5.013 \ \text{\AA}^3$.}
\tablenotetext[2]{4 oxygen and 2 nitrogen ligands, $\alpha(N^{3-})=2.684
\ \text{\AA}^3$.}
\tablenotetext[3]{5 oxygen ligands.} 
\tablenotetext[4]{5 oxygen and 2 nitrogen ligands.}
\end{table}

\begin{table}
\caption{Inter-shells potential, $V_{ss}$ ($\text{cm}^{-1}$), 
for a number of nearest oxygens $N$ between 8 and 12, 
${\bar {\text R}}=2.4\ \text{\AA}$, and for 
$\text {R}_{ss}=\text{R}_0=5.3\ \text{\AA}$.}
\label{tab2}
\begin{tabular}{cccccc}
 $N$ & 8 & 9 & 10 & 11 & 12 \\
\tableline
 $V_{ss}$ & 416.3 & 468.4 & 520.3 & 572.4 & 624.4 \\
\end{tabular}
\end{table}

\begin{references}
\bibitem[1]{Armand89}M. Armand and M. Gauthier, {\it High Conductivity
Solid Ionic Conductors} edited by T. Takahashi (World Scientific, 
Singapore, 1989), p. 114.

\bibitem[2]{Armand79}M. Armand, J.M. Chabagno and M. Duclot, {\it Fast 
Ion Transport in Solids}, edited by P. Vashishta, J.N. Mundy and G.K. 
Shenoy (North Holland, New York, 1979), p. 131.

\bibitem[3]{Chatani}Y. Chatani and S. Okamura, Polymer {\bf 28}, 1815
(1987); Y. Chatani, Y. Fujii, T. Takayanagi and A. Honma, Polymer 
{\bf 31}, 2238 (1990).

\bibitem[4]{Bruce}P. Lightfoot, M. A. Mehta and P.G. Bruce, Science 
{\bf 262}, 883 (1993); {\it idem}, J. Am. Chem. Soc. {\bf 116}, 7469  
(1994); P.G.Bruce, Electrochim. Acta {\bf 40}, 2077 (1995).

\bibitem[5]{Judd77}B.R. Judd, Phys. Rev. Lett. {\bf 39}, 242 (1977).

\bibitem[6]{Wright}P.V. Wright, {\it Polymer Electrolyte Reviews 2}, 
edited by J.R. MacCallum and C.A. Vincent (Elsevier Applied Science, 
London, 1989), p. 61.

\bibitem[7]{Carlos94b}L.D. Carlos and A.L.L. Videira, J. Chem. Phys. 
{\bf 101}, 8827 (1994).

\bibitem[8]{Carlos95a}L.D. Carlos, A.L.L. Videira, M. Assun\c c\~ao, 
and L. Alc\'acer, Electrochim. Acta {\bf 40}, 2143 (1995).

\bibitem[9]{Morrison80}C.A. Morrison, J. Chem. Phys. {\bf 72}, 1001
(1980); {\it idem}, {\it Angular Momentum Theory Applied to 
Interactions in Solids}, Lectures Notes in Chemistry Vol. 47 
(Springer-Verlag, Berlin, 1988).

\bibitem[10]{Carlos94a}L.D. Carlos and A.L.L. Videira, Phys. Rev. B. 
{\bf 49}, 11721 (1994).

\bibitem[11]{Carlos95b}L.D. Carlos, M. Assun\c c\~ao, and L. Alc\'acer,
J. Mater. Res. {\bf 10}, 202 (1995).

\bibitem[12]{Carlos95d}M.M.S. Puga, L.D. Carlos, T.M. Abrantes, and L.
Alc\'acer, Electrochim. Acta {\bf 40}, 2383 (1995); {\it idem}, Chem. 
Mater. {\bf 7}, 2316 (1995).

\bibitem[13]{Bernson}A. Bernson, J. Lindgren, W. Huang and R. Frech,
Polymer, {\bf 36}, 4471 (1995).

\bibitem[14]{Newman}D. J. Newman, Adv. Phys. {\bf 20}, 197 (1971).

\bibitem[15]{Leavitt}R.P. Leavitt, J. Chem. Phys. {\bf 77}, 1661 (1982).
\end{references}
\end{document}